\documentclass{moriond}

\def\Journal#1#2#3#4{{#1} {\bf #2}, #3 (#4)}

\def\PRD{{\em Phys.Rev.}D}

\def\be{\begin{equation}}
\def\ee{\end{equation}}
\def\bea{\begin{eqnarray}}
\def\eea{\end{eqnarray}}

\newcommand{\Photo}{\includegraphics[height=35mm]{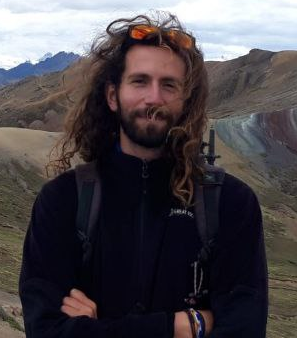}}

\begin{document}
\vspace*{4cm}
\title{Spontaneous ghostification:\\ how a dying black hole comes back as a naked singularity}

\author{ A. Bonanno${}^{1,2}$ and \underline{S. Silveravalle}${}^{3,4,5}$ }

\address{\mbox{${}^1$INAF, Osservatorio Astrofisico di Catania, Via S. Sofia 78, 95123 Catania, Italy}
\mbox{${}^2$INFN, Sezione di Catania, Via S. Sofia 64, 95123 Catania, Italy}
\mbox{${}^3$SISSA - International School for Advanced Studies, Via Bonomea 265, 34136 Trieste, Italy}
\mbox{${}^4$INFN, Sezione di Trieste, Via Valerio 2, 34127 Trieste, Italy}
\mbox{${}^5$IFPU - Institute for Fundamental Physics of the Universe, Via Beirut 2, 34151 Trieste, Italy}}

\maketitle\abstracts{A quantum ghost that destabilizes the Schwarzschild solution, transforming it into a naked singularity, may seem like a physicist’s worst nightmare. However, we argue that this scenario represents the natural evolution of a black hole within a conservative high-energy gravity framework and may, in fact, be a desirable outcome. Quadratic curvature terms typically appear as corrections to the Einstein-Hilbert action at high energies; nonetheless, such theories are generally considered incomplete due to the presence of ghost particles at the quantum level, which can spoil vacuum stability. We argue that this instability can only be triggered at the final stages of black hole evaporation, starting a phase transition-like process that alters the nature of the spacetime, similarly to spontaneous scalarization. We propose that the endpoint is a stable, exotic naked singularity, possible only in modified gravity theories, and avoids some of the pathological features associated with standard naked singularities.}

Our understanding of black holes was completely renewed by the derivation of their thermodynamic properties in the 1970s. Through the study of quantum fields in curved spacetime, Hawking showed that black holes can lose energy via the emission of particles. However, he soon realized that this process would ultimately lead to a breakdown in the theory’s predictability~\cite{sh1}. The formal origin of Hawking radiation lies in defining a vacuum state before collapse and observing the particle content after the event horizon forms. A heuristic interpretation is that, between the production and annihilation of virtual particle pairs, one particle falls into the black hole while the other escapes, extracting energy from the gravitational field. The emitted radiation thus carries information only about the external properties of the black hole, namely its mass and angular momentum, so that, by the end of evaporation, the knowledge on the microstate of the progenitor is irretrievably lost. However, the extreme energies present in the final stages of evaporation, where the black hole's temperature diverges, suggest that General Relativity should be modified by the inclusion of quantum corrections.

The form of the one-loop corrections to General Relativity inspired Stelle to study a theory in which all possible quadratic contractions of the curvature tensors are added to the Einstein-Hilbert action~\cite{ks1}. At the quantum level this theory, commonly referred to as quadratic gravity, is renormalizable, but at the cost of introducing a ghost particle with negative kinetic energy. While the presence of ghosts prevents quadratic gravity from being a well-defined fundamental theory of quantum gravity, it has nonetheless frequently appeared as a viable first-order correction to General Relativity. For this reason, we aim to investigate the possible evolution of a black hole in its final stages of evaporation, when quadratic corrections are taken into account.

In quadratic gravity, spherically symmetric black holes with sufficiently large masses (i.e. larger than $10^{-7}M_\odot$, though arguably even orders of magnitude smaller) are described only by the Schwarzschild metric. Astrophysical and primordial black holes are therefore expected to form with this geometry, and initially evaporate as predicted by Hawking. However, at a specific critical mass $M_c$ a transition to a non-Schwarzschild metric, characterized by a Yukawa-like correction to the gravitational potential~\cite{bs1}, is possible. An analysis of linear perturbations reveals that there exists a massive tensor mode, corresponding to the ghost of the quantum theory, that actually drives the transition to the non-Schwarzschild solution, with either an attractive or repulsive contribution from the Yukawa term~\cite{bs2,bs3}. The presence of long-lived, spatially extended perturbations at the critical mass suggests that this is indeed a phase transition in which the black hole acquires ghost Yukawa hair~\cite{bs3}. This process resembles spontaneous scalarization, where a scalar field instability triggered by strong gravity generates non-trivial hair.

The crucial aspect to understand is whether the Yukawa contribution to the potential, after the transition, is attractive or repulsive. In the former case, the total mass decreases (eventually becoming negative), and the horizon radius increases indefinitely; in the latter, the trend is reversed, and the horizon eventually shrinks to the origin with a finite, non-zero mass~\cite{bs1}. A fundamental difference lies in their linear stability: the large, Yukawa-attractive solutions are stable and in dynamical equilibrium, making the phase transition an equilibrium process; in contrast, the small, Yukawa-repulsive ones are unstable, leading to a non-equilibrium phase transition. Random fluctuations can perturb the critical solution toward either class of solutions, but while the former require the slow timescale of evaporation to deviate significantly from a Schwarzschild black hole, the latter exhibit an exponentially growing mode that quickly drives them away from it. The preferred direction for the evolution is therefore the unstable one~\cite{bs3}, which at the same time avoids the formation of large, gravitationally repulsive solutions~\cite{bs2}.

While a complete understanding of the evolution after the transition requires solving the non-linear, time-dependent equations, relevant physical insights can still be obtained through analytical approximations. The transition to a Yukawa-repulsive black hole changes the nature of the singularity, with all components of the metric vanishing as $r^2$ near the origin. This behavior remains stable during dynamical evolution, and we propose that the endpoint of this process is a static naked singularity with this near-origin geometry~\cite{bs3}. Preliminary analysis suggests that these solutions are linearly stable and exhibit infinite redshift for photons emitted at the singularity and measured at infinity~\cite{ss1}. In other words, only particles with infinite energy can escape the singularity, which would thus be hidden under a \emph{weaker} cosmic censorship conjecture.

Finally, the proposed mechanism effectively merges the information paradox with the singularity problem, leaving a remnant capable of storing the missing information.

\end{document}